\documentclass[%
reprint,
superscriptaddress,
showpacs,
 amsmath,amssymb,
 aps,
 prl,
]{revtex4-1}

\usepackage{graphicx}% Include figure files
\usepackage{dcolumn}% Align table columns on decimal point
\usepackage{bm}% bold math
\begin{document}

\preprint{APS/123-QED}

\title{Calorimetric study of multiaxial fluctuations in ferroelectric BaTiO$_3$ \protect\\ in nonpolar cubic phase}

\author{S. V. Grabovsky}
\email{grabovsky@physics.msu.ru}
\author{I. V. Shnaidshtein}
\affiliation{Faculty of Physics, Lomonosov Moscow State University, Moscow, 119991, Russia}
\author{M. Takesada}
\author{A. Onodera}
\affiliation{Department of Physics, Faculty of Science, Hokkaido University, Sapporo, 060-0810, Japan}
\author{B. A. Strukov}
\affiliation{Faculty of Physics, Lomonosov Moscow State University, Moscow, 119991, Russia}%

\date{\today}

\begin{abstract}
Precise specific heat of barium titanate single crystals of different quality has been measured with special attention to the temperature region above the ferroelectric phase transition. It is shown that excess specific heat of the multiaxial barium titanate in the paraelectric phase has a fluctuation nature and its temperature dependence is in accordance with the theoretical predictions of Levanyuk theory for multiaxial ferroelectrics. In this study the first observation of fluctuation contribution to the specific heat is presented. The correlation parameter $\delta$ is estimated to be 0.66$\times$10$^{-16}$\,cm$^2$ for Remeika-type crystals and 0.45$\times$10$^{-16}$\,cm$^2$ for TSSG (top-seeded solution growth) crystals.

%Keywords: ferroelectricity, barium titanate; specific heat; fluctuations
\end{abstract}

% PACS, the Physics and Astronomy
                             % Classification Scheme.
%\keywords{Suggested keywords}%Use showkeys class option if keyword
                              %display desired
\maketitle

%\tableofcontents

\section{Introduction}

Barium titanate~(BT) is a well-studied model ferroelectric crystal which has wide range of practical applications due to exceptional dielectric and electromechanical properties~\cite{Rabe}. However until now there are some unclear issues related to the mechanism of the ferroelectric phase transition. Recent works widely discuss the origin of the pretransitional phenomena in the cubic paraelectric phase which appear in the anomalous temperature dependence of Brillouin frequency shift~\cite{Ko2008,Ko2011}, birefringence~\cite{Ziebinska,Takagi}, thermal expansion~\cite{Rusek}, second harmonic generation~\cite{Pugachev}. Authors associate these effects with the hypothesis of the appearance of the local polar nanoregions (clusters) in the symmetric (on average) paraelectric phase~\cite{Zalar,Namikawa,Dulkin}. However the analytical form of the temperature dependence of the corresponding physical quantities was not determined. Only in ref.~\cite{Ko2012} possibility of a logarithmic temperature dependence of the elastic stiffness coefficient of BT near $T_c$ was noted.

Actually these effects can be the manifestation of the fluctuation phenomena, which can be considerable in the multiaxial ferroelectrics like BT~\cite{Strukov}. However the fluctuations of order parameter in BT have not been studied before in detail because the phase transitions in multiaxial ferroelectrics (what is BT) is of the first order. It is also essential to note that the fluctuation phenomena in uniaxial ferroelectrics (such as TGS or KDP) significantly suppressed and their experimental observation is impracticable~\cite{StrukovTGS}. This is true for both the order-disorder and displacive type of the ferroelectric phase transitions. Thus the study of fluctuation phenomena in multiaxial barium titanate becomes of fundamental interest, and this explains the noticeable activity in this direction. Calorimetric experiments can give significant contribution to the problem of anomalous pretransitional effects and fluctuation phenomena can give calorimetric experiments because historically traditional methods of determination of the specific heat are particularly accurate. This paper presents the results of such a precise calorimetric study of single crystals of BT of different quality with special attention to the temperature region above the ferroelectric phase transition point. Analysis includes accurate calculation of the anomalous part of the $C_p$, and revealing of its analytical form.

\section{Samples and Method}

The objects of the study were single crystals of barium titanate of various qualities. To clarify the defect influence we prepared single crystals grown by different methods. First group of the crystals were grown by Remeika method~\cite{Remeika} and represented yellow-colored triangle plates with the thickness of $0.2\div1.0$\,mm. The crystal samples of this type are generally considered as ``dirty'' ones containing excess of K, F, Pt~\cite{Godefroy}. The second group of the crystals was obtained by top-seeded solution growth method (TSSG)~\cite{Belruss}, they had linear dimensions of the order of $1$\,cm and were colorless. In each group, a few crystals were prepared for measurements. Crystal samples were cut to linear size of $2\div3$\,mm and then polished to a thickness of $0.10\div0.12$\,mm.

Specific heat measurements were carried out by {\it ac}\,-calorimetry technique using a calorimeter Shinku--Riko (ACC\,--1 M/L). The chopped light with the frequency of $2\div3$\,Hz from a halogen lamp was the source of thermal radiation. The method is based on the determination of amplitude of the temperature change of the sample under illumination of heat flux. Under certain conditions this change of the temperature is inversely proportional to the heat capacity of the sample. The method is described in details in~\cite{Sullivan,Hatta81}. Sensitivity of determination of specific heat~($C_p$) is $0.1\div0.2$\%. Since the {\it ac}\,-calorimetry method allows to obtain only relative values of specific heat we get absolute values of $C_p$ by comparison with the data of adiabatic calorimetry taken from the classical work of Todd~\cite{Todd}. Using this data we obtain that the specific heat at 302\,K is equal to 0.44\,J/g$\cdot$K.

The calorimetric experiments were performed by heating of the samples from room temperature to $500\div650$\,K and cooling back. The rate of temperature change was selected from 10 to 30\,K/h. The depth of the runs to the cubic paraelectric phase and time spent by sample at higher temperature were varied, so in fact it corresponded to the different conditions of the sample annealing. For the accurate determination of the background (lattice) heat capacity, the measurements were also made at low temperatures (down to 80\,K).

\section{Results and discussion}

It should be noted that we had qualitatively different thermal behaviour of the specific heat in the paraelectric phase for the first heating and cooling cycles. The example of temperature dependence of the specific heat is shown in Fig.~1 for the TSSG-type sample. It can be seen that the typical $\lambda$-type specific heat anomaly is observed in the vicinity of $T_c$ (406\,K), although the latent heat for first order phase transition was not detected in connection with the specific features of {\it ac}\,-calorimetry. Similar curves were obtained for the Remeika-type crystals in which $T_c$ was shifted down by $15\div20$\,K from $T_c$ for TSSG-type crystals (depending on crystal).
\begin{figure}[t]
\includegraphics[width=85mm]{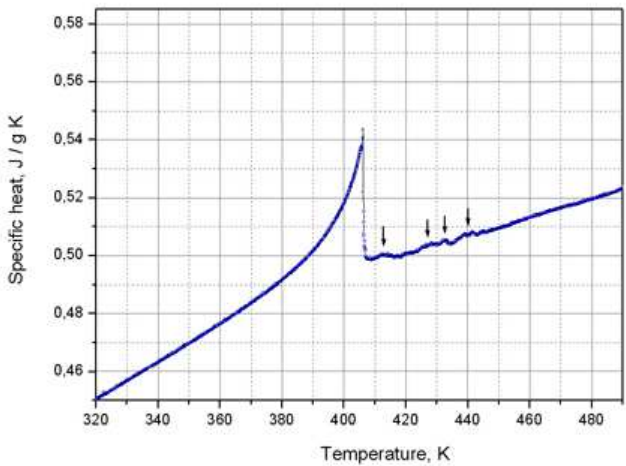}
\caption{Temperature dependence of the specific heat of BT single crystal (TSSG) at $T_c\pm100$\,K. Cooling.
Arrows show additional anomalies above $T_c$.}
\end{figure}

In the paraelectric phase above $T_c$, the additional diffused anomalies (marked by arrows) can be seen, which may remind us that observed previously by Hatta and Ikushima~\cite{Hatta76}. They only reported about an additional anomaly of specific heat at $T_c+20$\,K, but have not discussed it's nature in detail. We observed such anomalies on both Remeika and TSSG crystals, almost in all first measurements both on cooling or heating. In our experiments the additional anomalies appeared at different temperatures above $T_c$ and did not show a unified form.

But after several cycles of heating and cooling with deep run into the cubic paraelectric phase, we succeeded to obtain the monotonic dependences of the specific heat above $T_c$ without any additional anomalies both for Remeika and for TSSG crystals (Fig.~2). The appearance of additional anomalies and deviations of specific heat from monotonic curves should depend on the thermal history of the samples: this result clearly indicates the importance of pre-treatment of samples such as annealing, ageing and experimental conditions in an observation of such phenomena particularly near $T_c$.
\begin{figure}[b]
\includegraphics[width=85mm]{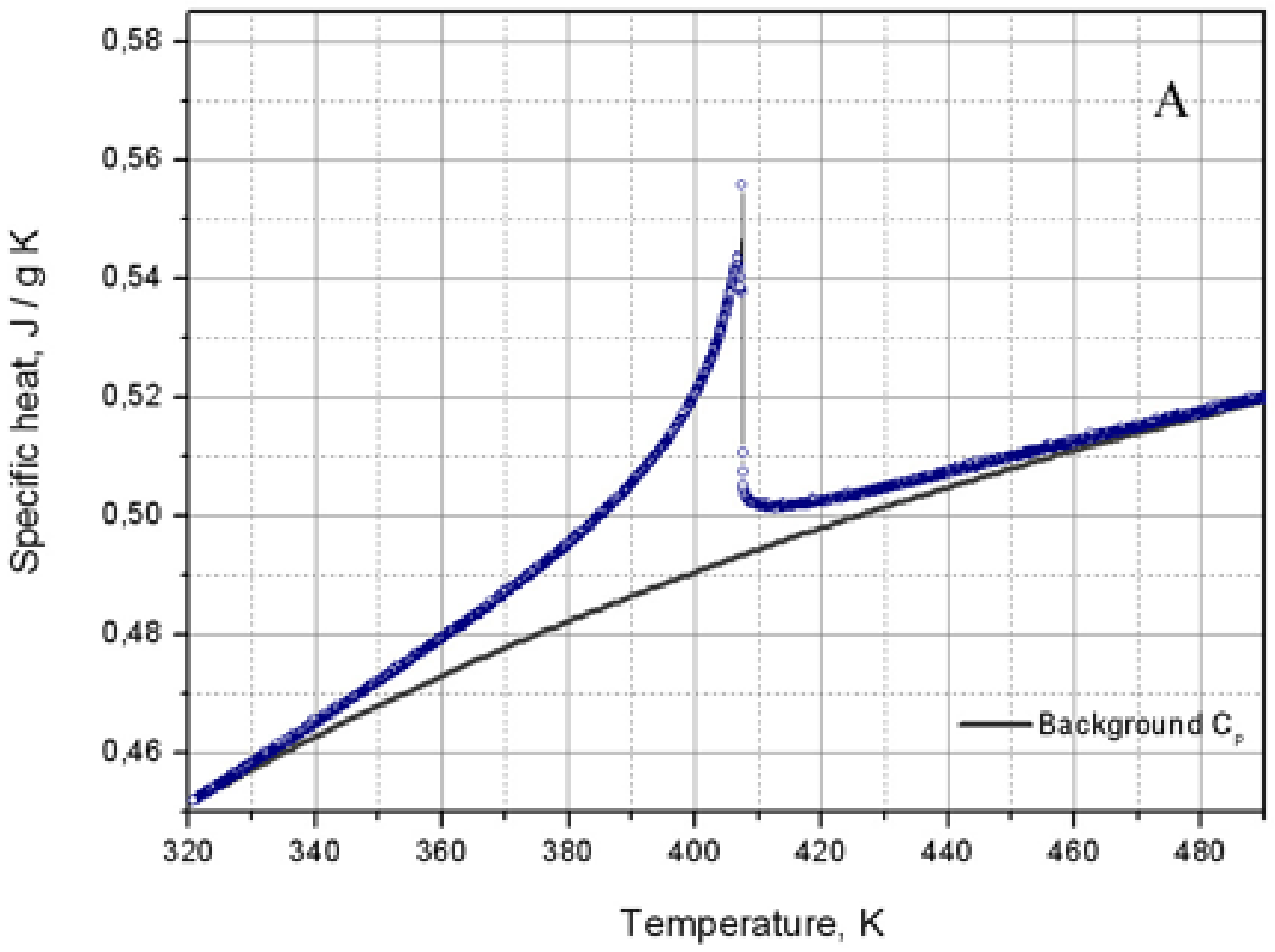}
\includegraphics[width=85mm]{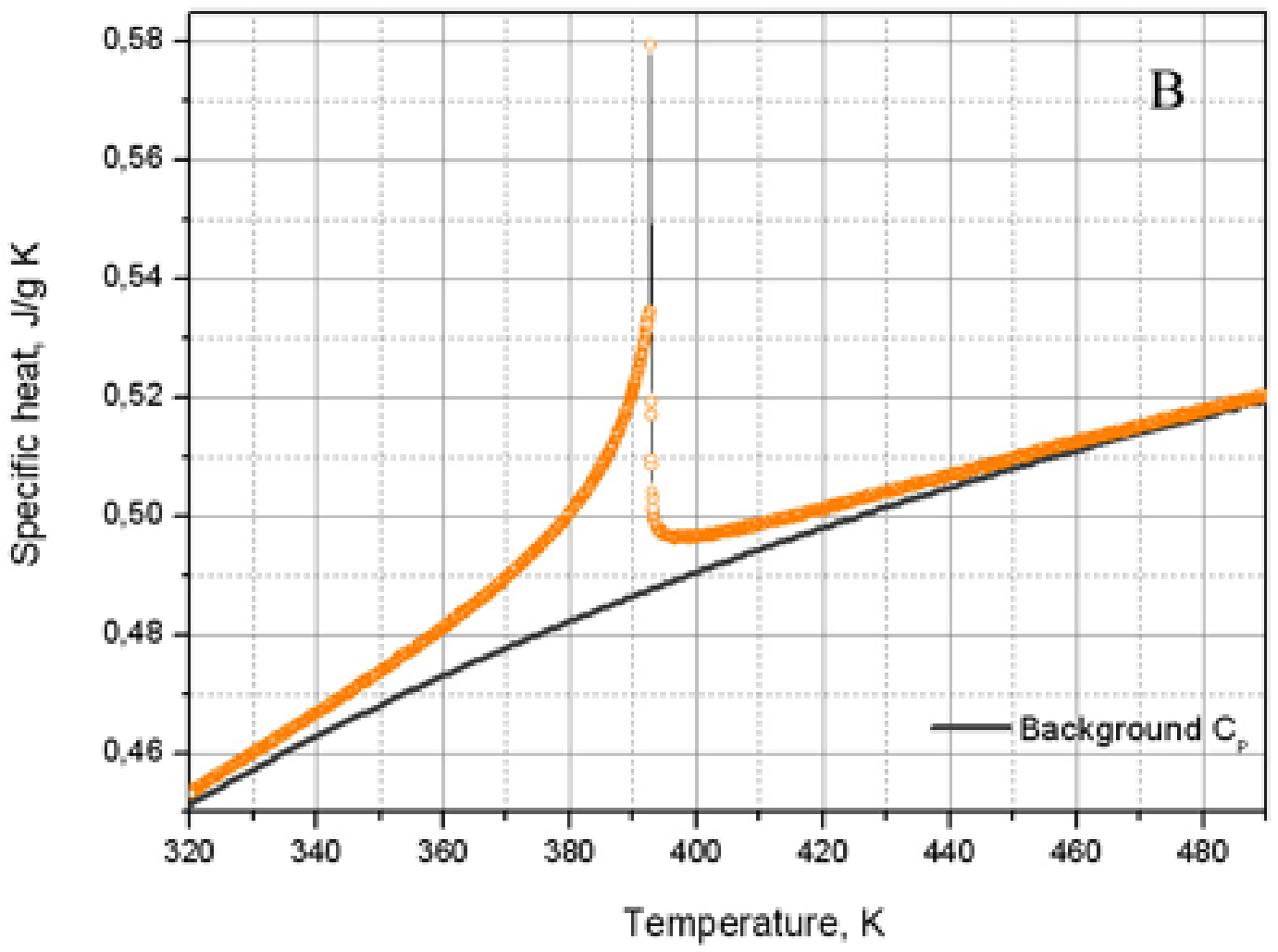}
\caption{Temperature dependence of the specific heat of BT single crystals at $T_c\pm100$\,K, after several heating-cooling runs. A -- TSSG, B -- Remeika-type crystal. Cooling. (Solid line: calculated background specific heat).}
\end{figure}

It is reasonable to assume that there is a kind of ``memory effect'' of the domain structure of the crystal, which persists even in the cubic phase. This memory effect is associated with the redistribution of defects, which leads to long relaxation processes after the transition to the paraelectric non-polar phase. When the defect reflection relaxes after thermal annealing during series of measurements, the associated thermal anomalies in the paraelectric phase disappeared (Fig.~2).

It is particularly important to point out that the same monotonic temperature dependence of $C_p$ above $T_c$ was observed for both groups of BT crystals. Thus it seems reasonable to consider the monotonic dependence of $C_p$ for the annealed BT crystals as their intrinsic property. For precise analysis, we have calculated the contribution of lattice part of $C_p$ (background) as
\begin{widetext}
\begin{equation}
C_p^{\mathrm{bg}}=\frac{3R}{\mu}\cdot\left[\mathcal{D}\left(\frac{477}T\right)+\mathcal{E}\left(\frac{194}T\right)+\mathcal{E}\left(\frac{416}T\right)+2\cdot\mathcal{E}\left(\frac{770}T\right)+7\times10^{-4}\cdot T\right]
\end{equation}
\end{widetext}
by using also low-temperature experimental data~\cite{Villar}, where $\mathcal{D}$ and $\mathcal{E}$ are Debye and Einstein functions, and $\mu=233.26$\,g/mol, $R$~--~universal gas constant.

The anomalous part of the specific heat ($C_p$) above this background for both types of BT crystals is clearly seen in Fig.~3, where $\Delta C_p(T)$ for TSSG and Remeika-type crystals are plotted against $T - T_c$. It is evident that there is a clear ``tail'' of the specific heat above $T_c$ in the temperature region between $T_c$ and $T_c+50$\,K and for the both types of crystal the curves almost coincide. That means that the effect of imperfections is excluded here and the intrinsic property of BT crystal is revealed.
\begin{figure}[t]
\includegraphics[width=85mm]{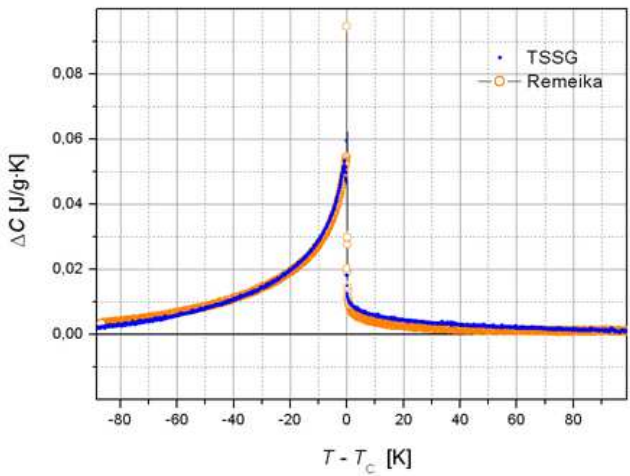}
\caption{Excess specific heat of BT for two types of crystals.}
\end{figure}

Figure~4 shows that $\Delta C_p/T^2$ is quite close to $(T - T_0)^{-1/2}$ behaviour in the rather wide temperature region above $T_c$. It seems to be a reasonable and important result because such temperature dependence was predicted by the theory of the fluctuation phenomena for the nonferroelectric and for the multiaxial ferroelectrics like BT~\cite{Levanyuk}:
\begin{equation}
\Delta C_p^{\mathrm{fl}}=\frac{k_{\mathrm{B}}T^2\alpha^{3/2}}{8\pi\delta^{3/2}}(T-T_0)^{-1/2}.
\end{equation}
Here $k_{\mathrm{B}}$ is Boltzmann constant, $\alpha=4\pi/C_{\mathrm{C-W}}$, and $C_{\mathrm{C-W}}$ is Curie--Weiss constant. The correlation parameter $\delta$ is an important physical quantity which determines the domain wall width and the polarization profile in thin films. It should be noted that in this formula we supposed to take not $T_c$ (transition temperature) but $T_0$ (extrapolated Curie--Weiss temperature for the first order phase transition). In BT, the difference $T_c - T_0$ is between 1\,K and 8\,K~\cite{Li2005}. The value 4\,K was chosen in our calculations to obtain the linear behaviour for $\Delta C_p^{\mathrm{fl}}$.
\begin{figure}[b]
\includegraphics[width=85mm]{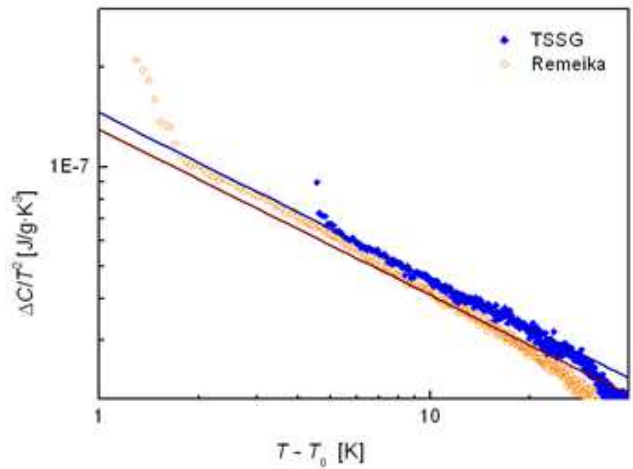}
\caption{Excess specific heat divided by $T^2$ for the two types of BT above phase transition in the double logarithmic scale.}
\end{figure}

Using the experimental data given in Fig.~4 we are able to estimate the coefficient of $(T - T_0)^{-1/2}$. As $C_{\mathrm{C-W}} = 2.07\times10^5$\,K for TSSG and $1.51\times10^5$\,K for Remeika-type crystal~\cite{Bednyakov}, the corresponding values of $\alpha$ are $6.07\times10^{-5}$\,K$^{-1}$ and $8.32\times10^{-5}$\,K$^{-1}$ respectively. Thus we obtained values of $\delta$ as $0.45\times10^{-16}$\,cm$^2$ for TSSG crystals and $0.66\times10^{-16}$\,cm$^2$ for Remeika crystals as shown in Fig.~5. It is interesting to note that the obtained optimal values of $\delta$ agree quite well with its estimation by the classical relation $\delta\sim\alpha T_c a^2$~\cite{Zhirnov}, where $a$ is the unit cell parameter (about $0.4$\,nm for BaTiO$_3$).
\begin{figure}[t]
\includegraphics[width=85mm]{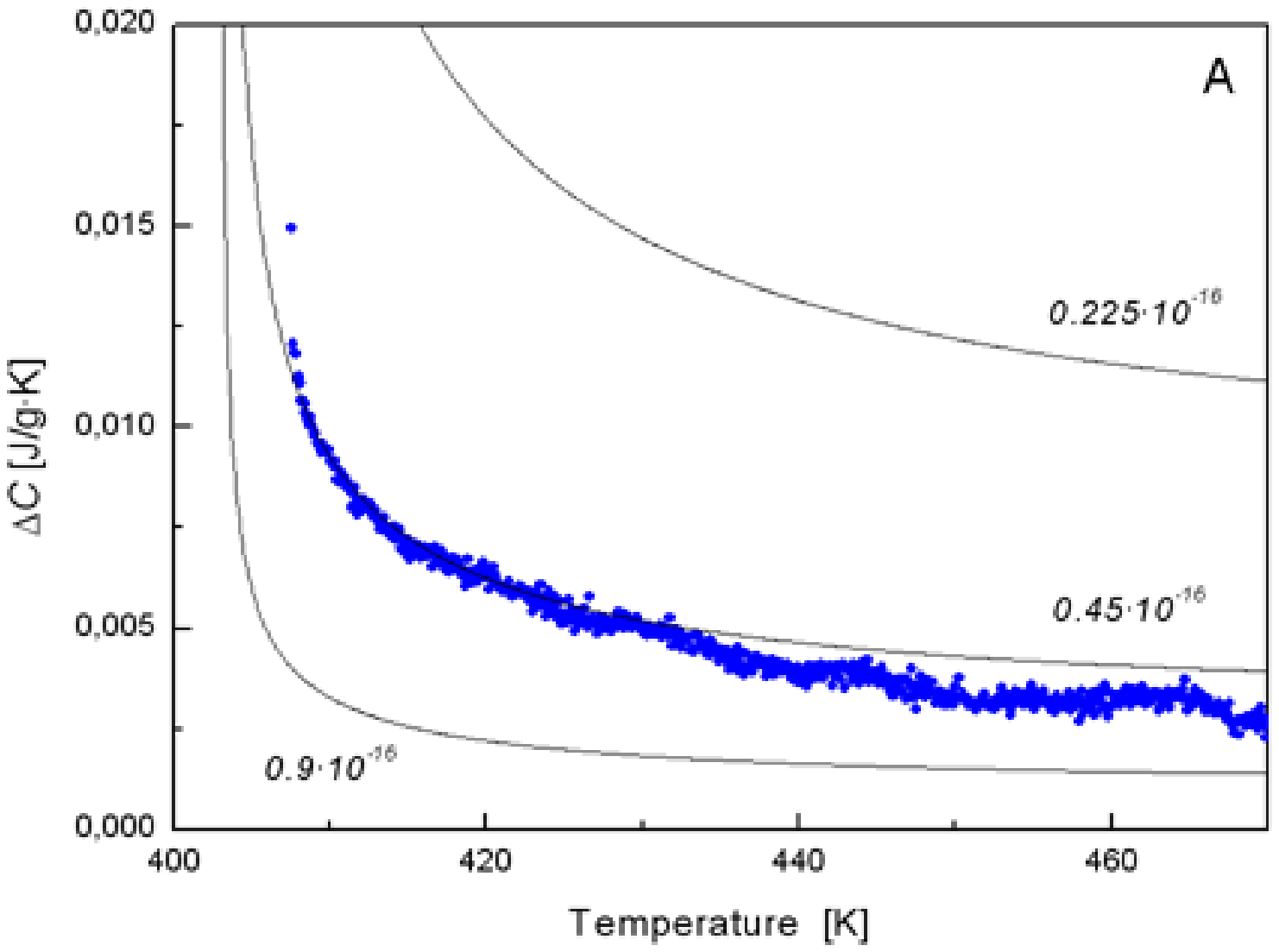}
\includegraphics[width=85mm]{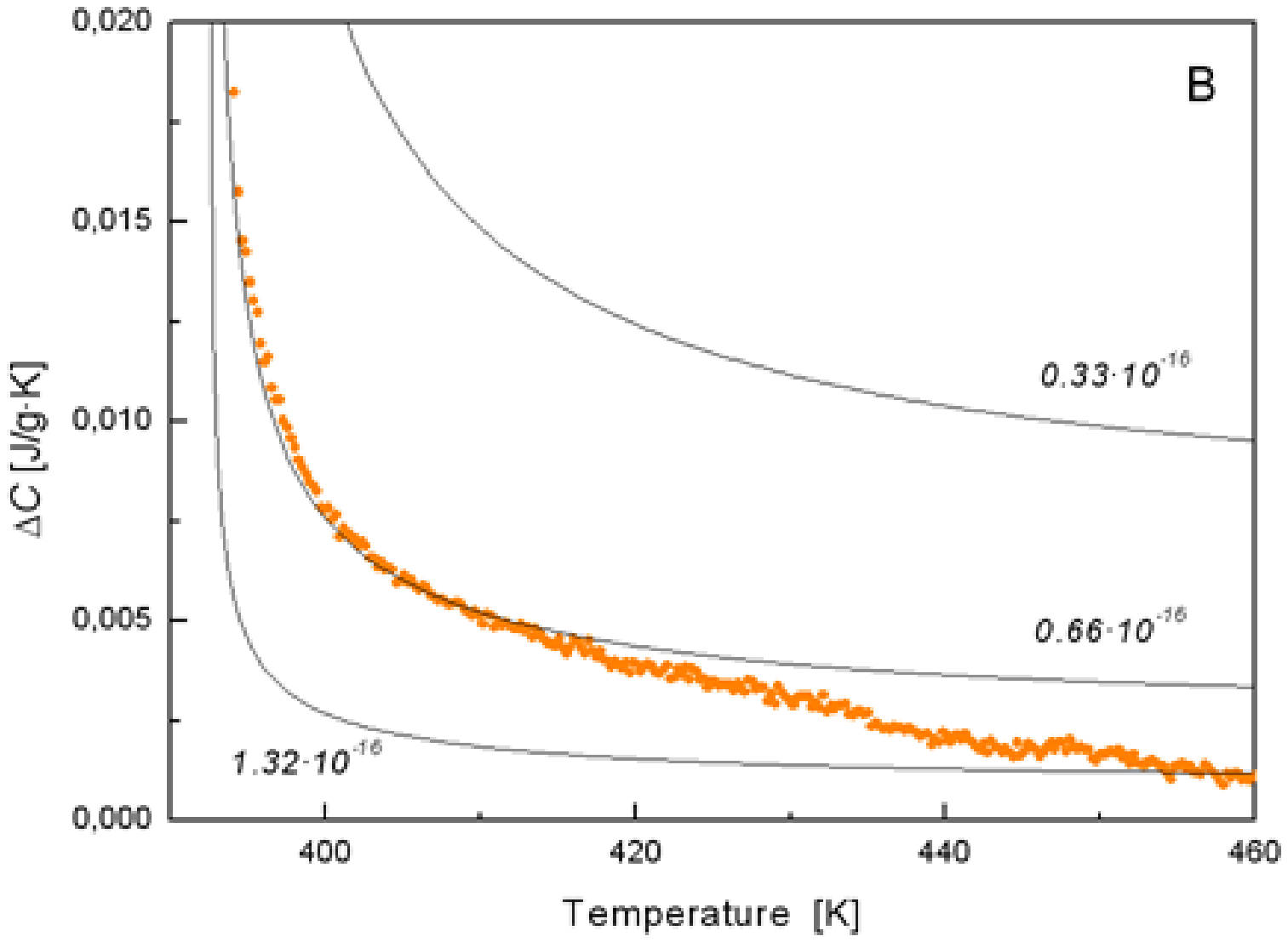}
\caption{Calculated temperature dependences of the fluctuation part of $C_p$ for three different values of $\delta$ (lines) and experimental temperature dependence of excess specific heat (dots). A -- TSSG, B -- Remeika-type crystal. Marked values of $\delta$ are in cm$^2$.}
\end{figure}

\section{Conclusion}

It has been already pointed that in uniaxial ferroelectrics of order-disorder type like TGS or KDP the polarization fluctuation contribution to specific heat is suppressed and quite small~\cite{Strukov,StrukovTGS}. In case of the multiaxial barium titanate, which cubic in the paraelectric phase, a situation is quite different. Fluctuation of polarization and the fluctuation contribution to the specific heat supposed to be considerable but still small, which has not been observed earlier because of the first order nature of the phase transition. In this study the first experimentally observed fluctuation contribution to the specific heat in the multiaxial ferroelectric crystal BT is presented. Described agreement between our data and Levanyuk's theory requires more detailed analysis, taking into account the behavior of the specific heat in the ferroelectric phase and the anisotropy of the $\delta_{ijkl}$ tensor, which will be published in the near future.

Using precise calorimetric measurements on the single-crystal barium titanate of different quality it is shown that the temperature dependence of specific heat in paraelectric phase can be obtained without additional anomalies above $T_c$ after multiple heating-cooling cycles. It is shown that ``tails'' of the specific heat in paraelectric phase has the fluctuation nature, the temperature dependence of anomalous part of specific heat in the cubic phase was found to be close to the inverse square root of $(T - T_0)$ that corresponds to theoretical values obtained by Levanyuk for multiaxial ferroelectric crystals. The effective value of the correlation parameter $\delta$ is successfully estimated.

\section{Acknowledgements}

The reported study was partially supported by RFBR, research project ¹11-02-01317-a.

\bibliography{grabovsky}

%merlin.mbs apsrev4-1.bst 2010-07-25 4.21a (PWD, AO, DPC) hacked
%Control: key (0)
%Control: author (8) initials jnrlst
%Control: editor formatted (1) identically to author
%Control: production of article title (-1) disabled
%Control: page (0) single
%Control: year (1) truncated
%Control: production of eprint (0) enabled
\providecommand{\noopsort}[1]{}\providecommand{\singleletter}[1]{#1}%
\begin{thebibliography}{25}%
\makeatletter
\providecommand \@ifxundefined [1]{%
 \@ifx{#1\undefined}
}%
\providecommand \@ifnum [1]{%
 \ifnum #1\expandafter \@firstoftwo
 \else \expandafter \@secondoftwo
 \fi
}%
\providecommand \@ifx [1]{%
 \ifx #1\expandafter \@firstoftwo
 \else \expandafter \@secondoftwo
 \fi
}%
\providecommand \natexlab [1]{#1}%
\providecommand \enquote  [1]{``#1''}%
\providecommand \bibnamefont  [1]{#1}%
\providecommand \bibfnamefont [1]{#1}%
\providecommand \citenamefont [1]{#1}%
\providecommand \href@noop [0]{\@secondoftwo}%
\providecommand \href [0]{\begingroup \@sanitize@url \@href}%
\providecommand \@href[1]{\@@startlink{#1}\@@href}%
\providecommand \@@href[1]{\endgroup#1\@@endlink}%
\providecommand \@sanitize@url [0]{\catcode `\\12\catcode `\$12\catcode
  `\&12\catcode `\#12\catcode `\^12\catcode `\_12\catcode `\%12\relax}%
\providecommand \@@startlink[1]{}%
\providecommand \@@endlink[0]{}%
\providecommand \url  [0]{\begingroup\@sanitize@url \@url }%
\providecommand \@url [1]{\endgroup\@href {#1}{\urlprefix }}%
\providecommand \urlprefix  [0]{URL }%
\providecommand \Eprint [0]{\href }%
\providecommand \doibase [0]{http://dx.doi.org/}%
\providecommand \selectlanguage [0]{\@gobble}%
\providecommand \bibinfo  [0]{\@secondoftwo}%
\providecommand \bibfield  [0]{\@secondoftwo}%
\providecommand \translation [1]{[#1]}%
\providecommand \BibitemOpen [0]{}%
\providecommand \bibitemStop [0]{}%
\providecommand \bibitemStop [0]{.\EOS\space}%
\providecommand \EOS [0]{\spacefactor3000\relax}%
\providecommand \BibitemShut  [1]{\csname bibitem#1\endcsname}%
\let\auto@bib@innerbib\@empty
%</preamble>
\bibitem [{\citenamefont {Rabe}\ \emph {et~al.}(2007)\citenamefont {Rabe},
  \citenamefont {Ahn},\ and\ \citenamefont {Triscone}}]{Rabe}%
  \BibitemOpen
  \bibinfo {editor} {\bibfnamefont {K.~M.}\ \bibnamefont {Rabe}}, \bibinfo
  {editor} {\bibfnamefont {C.~H.}\ \bibnamefont {Ahn}}, \ and\ \bibinfo
  {editor} {\bibfnamefont {J.-M.}\ \bibnamefont {Triscone}},\ eds.,\ \href@noop
  {} {\emph {\bibinfo {title} {Physics of Ferroelectrics: A Modern
  Perspective}}}\ (\bibinfo  {publisher} {Springer Publishing Company,
  Incorporated},\ \bibinfo {year} {2007})\BibitemShut {Stop}%
\bibitem [{\citenamefont {Ko}\ \emph {et~al.}(2011)\citenamefont {Ko},
  \citenamefont {Kim}, \citenamefont {Roleder}, \citenamefont {Rytz},\ and\
  \citenamefont {Kojima}}]{Ko2011}%
  \BibitemOpen
  \bibfield  {author} {\bibinfo {author} {\bibfnamefont {J.-H.}\ \bibnamefont
  {Ko}}, \bibinfo {author} {\bibfnamefont {T.~H.}\ \bibnamefont {Kim}},
  \bibinfo {author} {\bibfnamefont {K.}~\bibnamefont {Roleder}}, \bibinfo
  {author} {\bibfnamefont {D.}~\bibnamefont {Rytz}}, \ and\ \bibinfo {author}
  {\bibfnamefont {S.}~\bibnamefont {Kojima}},\ }\href@noop {} {\bibfield
  {journal} {\bibinfo  {journal} {Phys. Rev. B}\ }\textbf {\bibinfo {volume}
  {84}},\ \bibinfo {pages} {094123} (\bibinfo {year} {2011})}\BibitemShut
  {Stop}%
\bibitem [{\citenamefont {Ko}\ \emph {et~al.}(2008)\citenamefont {Ko},
  \citenamefont {Kojima}, \citenamefont {Koo}, \citenamefont {Jung},
  \citenamefont {Won},\ and\ \citenamefont {Hur}}]{Ko2008}%
  \BibitemOpen
  \bibfield  {author} {\bibinfo {author} {\bibfnamefont {J.-H.}\ \bibnamefont
  {Ko}}, \bibinfo {author} {\bibfnamefont {S.}~\bibnamefont {Kojima}}, \bibinfo
  {author} {\bibfnamefont {T.-Y.}\ \bibnamefont {Koo}}, \bibinfo {author}
  {\bibfnamefont {J.~H.}\ \bibnamefont {Jung}}, \bibinfo {author}
  {\bibfnamefont {C.~J.}\ \bibnamefont {Won}}, \ and\ \bibinfo {author}
  {\bibfnamefont {N.~J.}\ \bibnamefont {Hur}},\ }\href@noop {} {\bibfield
  {journal} {\bibinfo  {journal} {Appl. Phys. Lett.}\ }\textbf {\bibinfo
  {volume} {93}},\ \bibinfo {pages} {102905} (\bibinfo {year}
  {2008})}\BibitemShut {Stop}%
\bibitem [{\citenamefont {Ziebinska}\ \emph {et~al.}(2008)\citenamefont
  {Ziebinska}, \citenamefont {Rytz}, \citenamefont {Szot}, \citenamefont
  {Gorny},\ and\ \citenamefont {Roleder}}]{Ziebinska}%
  \BibitemOpen
  \bibfield  {author} {\bibinfo {author} {\bibfnamefont {A.}~\bibnamefont
  {Ziebinska}}, \bibinfo {author} {\bibfnamefont {D.}~\bibnamefont {Rytz}},
  \bibinfo {author} {\bibfnamefont {K.}~\bibnamefont {Szot}}, \bibinfo {author}
  {\bibfnamefont {M.}~\bibnamefont {Gorny}}, \ and\ \bibinfo {author}
  {\bibfnamefont {K.}~\bibnamefont {Roleder}},\ }\href@noop {} {\bibfield
  {journal} {\bibinfo  {journal} {Journal of Physics: Condensed Matter}\
  }\textbf {\bibinfo {volume} {20}},\ \bibinfo {pages} {142202} (\bibinfo
  {year} {2008})}\BibitemShut {Stop}%
\bibitem [{\citenamefont {Takagi}\ and\ \citenamefont
  {Ishidate}(2000)}]{Takagi}%
  \BibitemOpen
  \bibfield  {author} {\bibinfo {author} {\bibfnamefont {M.}~\bibnamefont
  {Takagi}}\ and\ \bibinfo {author} {\bibfnamefont {T.}~\bibnamefont
  {Ishidate}},\ }\href@noop {} {\bibfield  {journal} {\bibinfo  {journal}
  {Solid State Commun.}\ }\textbf {\bibinfo {volume} {113}},\ \bibinfo {pages}
  {423} (\bibinfo {year} {2000})}\BibitemShut {Stop}%
\bibitem [{\citenamefont {Rusek}\ \emph {et~al.}(2008)\citenamefont {Rusek},
  \citenamefont {Kruczek}, \citenamefont {Szot}, \citenamefont {Rytz},
  \citenamefont {Gorny},\ and\ \citenamefont {Roleder}}]{Rusek}%
  \BibitemOpen
  \bibfield  {author} {\bibinfo {author} {\bibfnamefont {K.}~\bibnamefont
  {Rusek}}, \bibinfo {author} {\bibfnamefont {J.}~\bibnamefont {Kruczek}},
  \bibinfo {author} {\bibfnamefont {K.}~\bibnamefont {Szot}}, \bibinfo {author}
  {\bibfnamefont {D.}~\bibnamefont {Rytz}}, \bibinfo {author} {\bibfnamefont
  {M.}~\bibnamefont {Gorny}}, \ and\ \bibinfo {author} {\bibfnamefont
  {K.}~\bibnamefont {Roleder}},\ }\href@noop {} {\bibfield  {journal} {\bibinfo
   {journal} {Ferroelectrics}\ }\textbf {\bibinfo {volume} {375}},\ \bibinfo
  {pages} {165} (\bibinfo {year} {2008})}\BibitemShut {Stop}%
\bibitem [{\citenamefont {Pugachev}\ \emph {et~al.}(2012)\citenamefont
  {Pugachev}, \citenamefont {Kovalevskii}, \citenamefont {Surovtsev},
  \citenamefont {Kojima}, \citenamefont {Prosandeev}, \citenamefont {Raevski},\
  and\ \citenamefont {Raevskaya}}]{Pugachev}%
  \BibitemOpen
  \bibfield  {author} {\bibinfo {author} {\bibfnamefont {A.~M.}\ \bibnamefont
  {Pugachev}}, \bibinfo {author} {\bibfnamefont {V.~I.}\ \bibnamefont
  {Kovalevskii}}, \bibinfo {author} {\bibfnamefont {N.~V.}\ \bibnamefont
  {Surovtsev}}, \bibinfo {author} {\bibfnamefont {S.}~\bibnamefont {Kojima}},
  \bibinfo {author} {\bibfnamefont {S.~A.}\ \bibnamefont {Prosandeev}},
  \bibinfo {author} {\bibfnamefont {I.~P.}\ \bibnamefont {Raevski}}, \ and\
  \bibinfo {author} {\bibfnamefont {S.~I.}\ \bibnamefont {Raevskaya}},\
  }\href@noop {} {\bibfield  {journal} {\bibinfo  {journal} {Phys. Rev. Lett.}\
  }\textbf {\bibinfo {volume} {108}},\ \bibinfo {pages} {247601} (\bibinfo
  {year} {2012})}\BibitemShut {Stop}%
\bibitem [{\citenamefont {Dul'kin}\ \emph {et~al.}(2010)\citenamefont
  {Dul'kin}, \citenamefont {Petzelt}, \citenamefont {Kamba}, \citenamefont
  {Mojaev},\ and\ \citenamefont {Roth}}]{Dulkin}%
  \BibitemOpen
  \bibfield  {author} {\bibinfo {author} {\bibfnamefont {E.}~\bibnamefont
  {Dul'kin}}, \bibinfo {author} {\bibfnamefont {J.}~\bibnamefont {Petzelt}},
  \bibinfo {author} {\bibfnamefont {S.}~\bibnamefont {Kamba}}, \bibinfo
  {author} {\bibfnamefont {E.}~\bibnamefont {Mojaev}}, \ and\ \bibinfo {author}
  {\bibfnamefont {M.}~\bibnamefont {Roth}},\ }\href@noop {} {\bibfield
  {journal} {\bibinfo  {journal} {Appl. Phys. Lett.}\ }\textbf {\bibinfo
  {volume} {97}},\ \bibinfo {pages} {032903} (\bibinfo {year}
  {2010})}\BibitemShut {Stop}%
\bibitem [{\citenamefont {Zalar}\ \emph {et~al.}(2005)\citenamefont {Zalar},
  \citenamefont {Lebar}, \citenamefont {Seliger}, \citenamefont {Blinc},
  \citenamefont {Laguta},\ and\ \citenamefont {Itoh}}]{Zalar}%
  \BibitemOpen
  \bibfield  {author} {\bibinfo {author} {\bibfnamefont {B.}~\bibnamefont
  {Zalar}}, \bibinfo {author} {\bibfnamefont {A.}~\bibnamefont {Lebar}},
  \bibinfo {author} {\bibfnamefont {J.}~\bibnamefont {Seliger}}, \bibinfo
  {author} {\bibfnamefont {R.}~\bibnamefont {Blinc}}, \bibinfo {author}
  {\bibfnamefont {V.~V.}\ \bibnamefont {Laguta}}, \ and\ \bibinfo {author}
  {\bibfnamefont {M.}~\bibnamefont {Itoh}},\ }\href@noop {} {\bibfield
  {journal} {\bibinfo  {journal} {Phys. Rev. B}\ }\textbf {\bibinfo {volume}
  {71}},\ \bibinfo {pages} {064107} (\bibinfo {year} {2005})}\BibitemShut
  {Stop}%
\bibitem [{\citenamefont {Namikawa}\ \emph {et~al.}(2009)\citenamefont
  {Namikawa}, \citenamefont {Kishimoto}, \citenamefont {Nasu}, \citenamefont
  {Matsushita}, \citenamefont {Tai}, \citenamefont {Sukegawa}, \citenamefont
  {Yamatani}, \citenamefont {Hasegawa}, \citenamefont {Nishikino},
  \citenamefont {Tanaka},\ and\ \citenamefont {Nagashima}}]{Namikawa}%
  \BibitemOpen
  \bibfield  {author} {\bibinfo {author} {\bibfnamefont {K.}~\bibnamefont
  {Namikawa}}, \bibinfo {author} {\bibfnamefont {M.}~\bibnamefont {Kishimoto}},
  \bibinfo {author} {\bibfnamefont {K.}~\bibnamefont {Nasu}}, \bibinfo {author}
  {\bibfnamefont {E.}~\bibnamefont {Matsushita}}, \bibinfo {author}
  {\bibfnamefont {R.~Z.}\ \bibnamefont {Tai}}, \bibinfo {author} {\bibfnamefont
  {K.}~\bibnamefont {Sukegawa}}, \bibinfo {author} {\bibfnamefont
  {H.}~\bibnamefont {Yamatani}}, \bibinfo {author} {\bibfnamefont
  {H.}~\bibnamefont {Hasegawa}}, \bibinfo {author} {\bibfnamefont
  {M.}~\bibnamefont {Nishikino}}, \bibinfo {author} {\bibfnamefont
  {M.}~\bibnamefont {Tanaka}}, \ and\ \bibinfo {author} {\bibfnamefont
  {K.}~\bibnamefont {Nagashima}},\ }\href@noop {} {\bibfield  {journal}
  {\bibinfo  {journal} {Phys. Rev. Lett.}\ }\textbf {\bibinfo {volume} {103}},\
  \bibinfo {pages} {197401} (\bibinfo {year} {2009})}\BibitemShut {Stop}%
\bibitem [{\citenamefont {Ko}\ \emph {et~al.}(2012)\citenamefont {Ko},
  \citenamefont {Kim}, \citenamefont {Kojima}, \citenamefont {Roleder},
  \citenamefont {Rytz}, \citenamefont {Won}, \citenamefont {Hur}, \citenamefont
  {Jung}, \citenamefont {Koo}, \citenamefont {Kim},\ and\ \citenamefont
  {Park}}]{Ko2012}%
  \BibitemOpen
  \bibfield  {author} {\bibinfo {author} {\bibfnamefont {J.-H.}\ \bibnamefont
  {Ko}}, \bibinfo {author} {\bibfnamefont {T.}~\bibnamefont {Kim}}, \bibinfo
  {author} {\bibfnamefont {S.}~\bibnamefont {Kojima}}, \bibinfo {author}
  {\bibfnamefont {K.}~\bibnamefont {Roleder}}, \bibinfo {author} {\bibfnamefont
  {D.}~\bibnamefont {Rytz}}, \bibinfo {author} {\bibfnamefont {C.}~\bibnamefont
  {Won}}, \bibinfo {author} {\bibfnamefont {N.}~\bibnamefont {Hur}}, \bibinfo
  {author} {\bibfnamefont {J.}~\bibnamefont {Jung}}, \bibinfo {author}
  {\bibfnamefont {T.-Y.}\ \bibnamefont {Koo}}, \bibinfo {author} {\bibfnamefont
  {S.}~\bibnamefont {Kim}}, \ and\ \bibinfo {author} {\bibfnamefont
  {K.}~\bibnamefont {Park}},\ }\href@noop {} {\bibfield  {journal} {\bibinfo
  {journal} {Current Applied Physics}\ }\textbf {\bibinfo {volume} {12}},\
  \bibinfo {pages} {1185} (\bibinfo {year} {2012})}\BibitemShut {Stop}%
\bibitem [{\citenamefont {Strukov}\ and\ \citenamefont
  {Levanyuk}(1998)}]{Strukov}%
  \BibitemOpen
  \bibfield  {author} {\bibinfo {author} {\bibfnamefont {B.~A.}\ \bibnamefont
  {Strukov}}\ and\ \bibinfo {author} {\bibfnamefont {A.~P.}\ \bibnamefont
  {Levanyuk}},\ }\href@noop {} {\emph {\bibinfo {title} {Ferroelectric
  Phenomena in Crystals}}}\ (\bibinfo  {publisher} {Springer Publishing
  Company, Incorporated},\ \bibinfo {year} {1998})\BibitemShut {Stop}%
\bibitem [{\citenamefont {Strukov}\ \emph {et~al.}(1998)\citenamefont
  {Strukov}, \citenamefont {Ragula}, \citenamefont {Arkhangel'skaya},\ and\
  \citenamefont {Shnaidshtein}}]{StrukovTGS}%
  \BibitemOpen
  \bibfield  {author} {\bibinfo {author} {\bibfnamefont {B.}~\bibnamefont
  {Strukov}}, \bibinfo {author} {\bibfnamefont {E.}~\bibnamefont {Ragula}},
  \bibinfo {author} {\bibfnamefont {S.}~\bibnamefont {Arkhangel'skaya}}, \ and\
  \bibinfo {author} {\bibfnamefont {I.}~\bibnamefont {Shnaidshtein}},\
  }\href@noop {} {\bibfield  {journal} {\bibinfo  {journal} {Physics of the
  Solid State}\ }\textbf {\bibinfo {volume} {40}},\ \bibinfo {pages} {94}
  (\bibinfo {year} {1998})}\BibitemShut {Stop}%
\bibitem [{\citenamefont {Remeika}\ and\ \citenamefont
  {Jackson}(1954)}]{Remeika}%
  \BibitemOpen
  \bibfield  {author} {\bibinfo {author} {\bibfnamefont {J.~P.}\ \bibnamefont
  {Remeika}}\ and\ \bibinfo {author} {\bibfnamefont {W.~M.}\ \bibnamefont
  {Jackson}},\ }\href@noop {} {\bibfield  {journal} {\bibinfo  {journal}
  {Journal of the American Chemical Society}\ }\textbf {\bibinfo {volume}
  {76}},\ \bibinfo {pages} {940} (\bibinfo {year} {1954})}\BibitemShut
  {Stop}%
\bibitem [{\citenamefont {Godefroy}\ \emph {et~al.}(1977)\citenamefont
  {Godefroy}, \citenamefont {Lompre}, \citenamefont {Dumas},\ and\
  \citenamefont {Arend}}]{Godefroy}%
  \BibitemOpen
  \bibfield  {author} {\bibinfo {author} {\bibfnamefont {G.}~\bibnamefont
  {Godefroy}}, \bibinfo {author} {\bibfnamefont {P.}~\bibnamefont {Lompre}},
  \bibinfo {author} {\bibfnamefont {C.}~\bibnamefont {Dumas}}, \ and\ \bibinfo
  {author} {\bibfnamefont {H.}~\bibnamefont {Arend}},\ }\href@noop {}
  {\bibfield  {journal} {\bibinfo  {journal} {Materials Research Bulletin}\
  }\textbf {\bibinfo {volume} {12}},\ \bibinfo {pages} {165} (\bibinfo {year}
  {1977})}\BibitemShut {Stop}%
\bibitem [{\citenamefont {Belruss}\ \emph {et~al.}(1971)\citenamefont
  {Belruss}, \citenamefont {Kalnajs}, \citenamefont {Linz},\ and\ \citenamefont
  {Folweiler}}]{Belruss}%
  \BibitemOpen
  \bibfield  {author} {\bibinfo {author} {\bibfnamefont {V.}~\bibnamefont
  {Belruss}}, \bibinfo {author} {\bibfnamefont {J.}~\bibnamefont {Kalnajs}},
  \bibinfo {author} {\bibfnamefont {A.}~\bibnamefont {Linz}}, \ and\ \bibinfo
  {author} {\bibfnamefont {R.}~\bibnamefont {Folweiler}},\ }\href@noop {}
  {\bibfield  {journal} {\bibinfo  {journal} {Materials Research Bulletin}\
  }\textbf {\bibinfo {volume} {6}},\ \bibinfo {pages} {899} (\bibinfo {year}
  {1971})}\BibitemShut {Stop}%
\bibitem [{\citenamefont {Sullivan}\ and\ \citenamefont
  {Seidel}(1968)}]{Sullivan}%
  \BibitemOpen
  \bibfield  {author} {\bibinfo {author} {\bibfnamefont {P.~F.}\ \bibnamefont
  {Sullivan}}\ and\ \bibinfo {author} {\bibfnamefont {G.}~\bibnamefont
  {Seidel}},\ }\href@noop {} {\bibfield  {journal} {\bibinfo  {journal} {Phys.
  Rev.}\ }\textbf {\bibinfo {volume} {173}},\ \bibinfo {pages} {679} (\bibinfo
  {year} {1968})}\BibitemShut {Stop}%
\bibitem [{\citenamefont {Hatta}\ and\ \citenamefont
  {Ikushima}(1981)}]{Hatta81}%
  \BibitemOpen
  \bibfield  {author} {\bibinfo {author} {\bibfnamefont {I.}~\bibnamefont
  {Hatta}}\ and\ \bibinfo {author} {\bibfnamefont {A.~J.}\ \bibnamefont
  {Ikushima}},\ }\href@noop {} {\bibfield  {journal} {\bibinfo  {journal}
  {Japanese Journal of Applied Physics}\ }\textbf {\bibinfo {volume} {20}},\
  \bibinfo {pages} {1995} (\bibinfo {year} {1981})}\BibitemShut {Stop}%
\bibitem [{\citenamefont {Todd}\ and\ \citenamefont {Lorenson}(1952)}]{Todd}%
  \BibitemOpen
  \bibfield  {author} {\bibinfo {author} {\bibfnamefont {S.~S.}\ \bibnamefont
  {Todd}}\ and\ \bibinfo {author} {\bibfnamefont {R.~E.}\ \bibnamefont
  {Lorenson}},\ }\href@noop {} {\bibfield  {journal} {\bibinfo  {journal}
  {Journal of the American Chemical Society}\ }\textbf {\bibinfo {volume}
  {74}},\ \bibinfo {pages} {2043} (\bibinfo {year} {1952})}\BibitemShut
  {Stop}%
\bibitem [{\citenamefont {Hatta}\ and\ \citenamefont
  {Ikushima}(1976)}]{Hatta76}%
  \BibitemOpen
  \bibfield  {author} {\bibinfo {author} {\bibfnamefont {I.}~\bibnamefont
  {Hatta}}\ and\ \bibinfo {author} {\bibfnamefont {A.}~\bibnamefont
  {Ikushima}},\ }\href@noop {} {\bibfield  {journal} {\bibinfo  {journal}
  {Journal of the Physical Society of Japan}\ }\textbf {\bibinfo {volume}
  {41}},\ \bibinfo {pages} {558} (\bibinfo {year} {1976})}\BibitemShut
  {Stop}%
\bibitem [{\citenamefont {Villar}\ \emph {et~al.}(1986)\citenamefont {Villar},
  \citenamefont {Gmelin},\ and\ \citenamefont {Grimm}}]{Villar}%
  \BibitemOpen
  \bibfield  {author} {\bibinfo {author} {\bibfnamefont {R.}~\bibnamefont
  {Villar}}, \bibinfo {author} {\bibfnamefont {E.}~\bibnamefont {Gmelin}}, \
  and\ \bibinfo {author} {\bibfnamefont {H.}~\bibnamefont {Grimm}},\
  }\href@noop {} {\bibfield  {journal} {\bibinfo  {journal} {Ferroelectrics}\
  }\textbf {\bibinfo {volume} {69}},\ \bibinfo {pages} {165} (\bibinfo {year}
  {1986})}\BibitemShut {Stop}%
\bibitem [{\citenamefont {Levanyuk}(1964)}]{Levanyuk}%
  \BibitemOpen
  \bibfield  {author} {\bibinfo {author} {\bibfnamefont {A.}~\bibnamefont
  {Levanyuk}},\ }\href@noop {} {\bibfield  {journal} {\bibinfo  {journal} {Sov.
  Phys. Solid State}\ }\textbf {\bibinfo {volume} {5}},\ \bibinfo {pages}
  {1294} (\bibinfo {year} {1964})}\BibitemShut {Stop}%
\bibitem [{\citenamefont {Li}\ \emph {et~al.}(2005)\citenamefont {Li},
  \citenamefont {Cross},\ and\ \citenamefont {Chen}}]{Li2005}%
  \BibitemOpen
  \bibfield  {author} {\bibinfo {author} {\bibfnamefont {Y.~L.}\ \bibnamefont
  {Li}}, \bibinfo {author} {\bibfnamefont {L.~E.}\ \bibnamefont {Cross}}, \
  and\ \bibinfo {author} {\bibfnamefont {L.~Q.}\ \bibnamefont {Chen}},\
  }\href@noop {} {\bibfield  {journal} {\bibinfo  {journal} {Journal of Applied
  Physics}\ }\textbf {\bibinfo {volume} {98}},\ \bibinfo {pages} {064101}
  (\bibinfo {year} {2005})}\BibitemShut {Stop}%
\bibitem [{\citenamefont {Bednyakov}\ \emph {et~al.}(2011)\citenamefont
  {Bednyakov}, \citenamefont {Shnaidshtein},\ and\ \citenamefont
  {Strukov}}]{Bednyakov}%
  \BibitemOpen
  \bibfield  {author} {\bibinfo {author} {\bibfnamefont {P.}~\bibnamefont
  {Bednyakov}}, \bibinfo {author} {\bibfnamefont {I.}~\bibnamefont
  {Shnaidshtein}}, \ and\ \bibinfo {author} {\bibfnamefont {B.}~\bibnamefont
  {Strukov}},\ }\href@noop {} {\bibfield  {journal} {\bibinfo  {journal}
  {Physics of the Solid State}\ }\textbf {\bibinfo {volume} {53}},\ \bibinfo
  {pages} {350} (\bibinfo {year} {2011})}\BibitemShut {Stop}%
\bibitem [{\citenamefont {Zhirnov}(1959)}]{Zhirnov}%
  \BibitemOpen
  \bibfield  {author} {\bibinfo {author} {\bibfnamefont {V.}~\bibnamefont
  {Zhirnov}},\ }\href@noop {} {\bibfield  {journal} {\bibinfo  {journal} {Sov.
  Phys. JETP}\ }\textbf {\bibinfo {volume} {35}},\ \bibinfo {pages} {822}
  (\bibinfo {year} {1959})}\BibitemShut {Stop}%
\end{thebibliography}%

\end{document}